# USAGE OF THE SYMBOLIC METHOD AND EXACT SOLUTIONS OF HARMONIC FUNCTIONS IN THE PROBLEM ON PRESSING OF THE FLAT STAMP IN THE RECTANGULAR PLATE

Sabo I.I., Tolok V.O.

Zaporizhzhya National University, Ukraine

## ABSTRACT

In this paper we construct a solution of the two-dimensional problem of the elasticity theory (plane deformation) on pressing of the flat stamp in the rectangular plate by using the symbolic method and the exact solution of the harmonic equation for a rectangle. The substitution of symbolic functions that satisfy the harmonic equation, with relevant solutions of harmonic functions allows obtaining the exact solution of the problem.

## INTRODUCTION

The symbolic solution for the strip can be obtained from the symbolic solution of Vlasov V.Z. for initial functions, which is the sum of the multiplications of differential operators and the corresponding initial functions [1]:

$$U(x,y) = Gu(x,y) = L_{UU}U_0(x) + L_{UV}V_0(x) + L_{UY}Y_0(x) + L_{UX}X_0(x),$$

$$V(x,y) = Gv(x,y) = L_{VU}U_0(x) + L_{VV}V_0(x) + L_{VY}Y_0(x) + L_{VX}X_0(x),$$

$$Y(x,y) = \sigma_y(x,y) = L_{YU}U_0(x) + L_{YV}V_0(x) + L_{YY}Y_0(x) + L_{YX}X_0(x),$$

$$X(x,y) = \tau_{xy}(x,y) = L_{XU}U_0(x) + L_{XV}V_0(x) + L_{XY}Y_0(x) + L_{XX}X_0(x),$$

$$\sigma_x(x,y) = A_U U_0(x) + A_V V_0(x) + A_Y Y_0(x) + A_X X_0(x),$$

where $U_0(x) = Gu(x,0)$, $V_0(x) = Gv(x,0)$, $Y_0(x) = \sigma_y(x,0)$, $X_0(x) = \tau_{xy}(x,0)$ – initial functions.

Differential operators can be represented either in the form of infinite operating series, or in the form of symbolic formulas (for a plane-deformed state) [1]:

$$L_{UU} = \cos(\alpha y) - \frac{\alpha y}{2(1-\nu)}\sin(\alpha y) \quad L_{UV} = -\frac{(1-2\nu)}{2(1-\nu)}\sin(\alpha y) - \frac{\alpha y}{2(1-\nu)}\cos(\alpha y),$$

$$L_{UY} = -\frac{y}{4(1-\nu)}\sin(\alpha y) \quad L_{UX} = \frac{1}{\alpha}\sin(\alpha y) - \frac{\sin(\alpha y)}{4\alpha(1-\nu)} - \frac{\alpha y\cos(\alpha y)}{4\alpha(1-\nu)},$$

$$L_{VU} = \frac{1-2\nu}{2(1-\nu)}\sin(\alpha y) - \frac{\alpha y}{2(1-\nu)}\cos(\alpha y) \quad L_{VV} = \frac{\alpha y}{2(1-\nu)}\sin(\alpha y) + \cos(\alpha y),$$

$$L_{YY} = \frac{(3-4\nu)\sin(\alpha y)}{4(1-\nu)\alpha} - \frac{y}{4(1-\nu)}\cos(\alpha y) \quad, \quad L_{VX} = L_{UY} \quad,$$

$$L_{YU} = \frac{\alpha^2 y}{1-\nu}\sin(\alpha y) \quad, \quad L_{YV} = \frac{\alpha}{1-\nu}(\sin(\alpha y) - \alpha y \cos(\alpha y)) \quad,$$

$$L_{YY} = L_{VV} \quad, \quad L_{YX} = L_{UV} \quad,$$

$$L_{XU} = -\frac{\alpha}{1-\nu}(\sin(\alpha y) + \alpha y \cos(\alpha y)) \quad, \quad L_{XV} = L_{YU} \quad,$$

$$L_{XY} = L_{VU} \quad, \quad L_{XX} = L_{UU} \quad,$$

$$A_U = \frac{2\alpha}{1-\nu}\cos(\alpha y) - \frac{y\alpha^2}{1-\nu}\sin(\alpha y) \quad, \quad A_V = -\frac{\alpha}{1-\nu}(\sin(\alpha y) + \alpha y \cos(\alpha y)) \quad,$$

$$A_Y = \frac{\nu}{1-\nu}\cos(\alpha y) - \frac{y\alpha}{2(1-\nu)}\sin(\alpha y) \quad, \quad A_X = \frac{y\alpha}{2(1-\nu)}\cos(\alpha y) + \frac{3-2\nu}{2(1-\nu)}\sin(\alpha y) \quad,$$

where $\alpha = \dfrac{\partial}{\partial x}$ .

In this paper differential operators is replaced by harmonic solutions in the symbolic solution for the strip. This allows us to get an exact solution of the problem.

## 1. SYMBOLIC SOLUTION FOR THE STRIPE

Let us find a symbolic solution for the strip that satisfies the following boundary conditions:

$$V(x,0) = V_0(x) = 0 \quad, \tag{1}$$

$$X(x,0) = X_0(x) = 0 \quad, \tag{2}$$

$$V(x,h) = V_h(x) \quad, \tag{3}$$

$$X(x,h) = X_h(x) = 0 \quad. \tag{4}$$

To get a symbolic solution we substitute the symbolic solution of Vlasov V.Z. in the boundary conditions (1) – (4). Solving the resulting system of four equations relatively the four initial functions, we obtain an expression for all four initial functions $U_0(x)$, $V_0(x)$, $Y_0(x)$, $X_0(x)$ through the function on the boundary $V_h(x)$. Substituting these expressions into a symbolic solution for the initial functions, we get a symbolic solution for the strip satisfying the boundary conditions (1) – (4):

$$U(x,y) = \frac{1+\nu}{2}\left( y\frac{\alpha\sin(y\alpha)}{\sin(h\alpha)} - \left(\frac{1-\nu}{1+\nu}\right)\frac{\cos(y\alpha)}{\sin(h\alpha)} + h\frac{\alpha\cos(h\alpha)\cos(y\alpha)}{(\sin(h\alpha))^2} \right)V_h(x) \quad, \tag{5}$$

$$V(x,y) = \frac{1+\nu}{2}\left(\left(\frac{2}{1+\nu}\right)\frac{\sin(y\alpha)}{\sin(h\alpha)} - y\frac{\alpha\cos(y\alpha)}{\sin(h\alpha)} + h\frac{\alpha\cos(h\alpha)\sin(y\alpha)}{(\sin(h\alpha))^2}\right)V_h(x) \tag{6}$$

$$Y(x,y) = \frac{E}{2}\left(\frac{\alpha\cos(y\alpha)}{\sin(h\alpha)} + y\frac{\alpha^2\sin(y\alpha)}{\sin(h\alpha)} + h\frac{\alpha^2\cos(h\alpha)\cos(y\alpha)}{(\sin(h\alpha))^2}\right)V_h(x) \tag{7}$$

$$X(x,y) = \frac{E}{2}\left(h\frac{\alpha^2\cos(h\alpha)\sin(y\alpha)}{(\sin(h\alpha))^2} - y\frac{\alpha^2\cos(y\alpha)}{\sin(h\alpha)}\right)V_h(x) \tag{8}$$

$$\sigma_x(x,y) = \frac{E}{2}\left(\frac{\alpha\cos(y\alpha)}{\sin(h\alpha)} - y\frac{\alpha^2\sin(y\alpha)}{\sin(h\alpha)} - h\frac{\alpha^2\cos(h\alpha)\cos(y\alpha)}{(\sin(h\alpha))^2}\right)V_h(x) \tag{9}$$

The obtained relations satisfy the equations of equilibrium, the equations expressing the relationship between stresses and displacements, and also the biharmonic equation.

## 2. EXACT SOLUTION OF HARMONIC FUNCTIONS

The resulting symbolic solution for the strip (5) - (9) contains the following harmonic functions [2]:

$$\frac{\sin(y\alpha)}{\sin(h\alpha)}V_h(x) \tag{10}$$

$$\frac{\cos(y\alpha)}{\sin(h\alpha)}V_h(x) = \frac{1}{\alpha}\frac{\partial}{\partial y}\left(\frac{\sin(y\alpha)}{\sin(h\alpha)}V_h(x)\right) = \int\frac{\partial}{\partial y}\left(\frac{\sin(y\alpha)}{\sin(h\alpha)}V_h(x)\right)dx \tag{11}$$

$$\frac{\alpha\sin(y\alpha)}{\sin(h\alpha)}V_h(x) = \alpha\left(\frac{\sin(y\alpha)}{\sin(h\alpha)}V_h(x)\right) = \frac{\partial}{\partial x}\left(\frac{\sin(y\alpha)}{\sin(h\alpha)}V_h(x)\right) \tag{12}$$

$$\frac{\alpha\cos(y\alpha)}{\sin(h\alpha)}V_h(x) = \frac{\partial}{\partial y}\left(\frac{\sin(y\alpha)}{\sin(h\alpha)}V_h(x)\right) \tag{13}$$

$$\frac{\alpha^2\sin(y\alpha)}{\sin(h\alpha)}V_h(x) = \alpha^2\left(\frac{\sin(y\alpha)}{\sin(h\alpha)}V_h(x)\right) = \frac{\partial^2}{\partial x^2}\left(\frac{\sin(y\alpha)}{\sin(h\alpha)}V_h(x)\right) \tag{14}$$

$$\frac{\alpha^2\cos(y\alpha)}{\sin(h\alpha)}V_h(x) = \alpha\frac{\partial}{\partial y}\left(\frac{\sin(y\alpha)}{\sin(h\alpha)}V_h(x)\right) = \frac{\partial^2}{\partial x\partial y}\left(\frac{\sin(y\alpha)}{\sin(h\alpha)}V_h(x)\right) \tag{15}$$

$$\frac{\alpha\cos(h\alpha)\cos(y\alpha)}{(\sin(h\alpha))^2}V_h(x) = -\frac{\partial}{\partial h}\left(\frac{\cos(y\alpha)}{\sin(h\alpha)}V_h(x)\right) = -\int\frac{\partial^2}{\partial h\partial y}\left(\frac{\sin(y\alpha)}{\sin(h\alpha)}V_h(x)\right)dx \tag{16}$$

$$\frac{\alpha^2\cos(h\alpha)\cos(y\alpha)}{(\sin(h\alpha))^2}V_h(x) = -\alpha\frac{\partial}{\partial h}\left(\frac{\cos(y\alpha)}{\sin(h\alpha)}V_h(x)\right) = -\frac{\partial^2}{\partial h\partial y}\left(\frac{\sin(y\alpha)}{\sin(h\alpha)}V_h(x)\right) \tag{17}$$

It is known that if a harmonic function $f(x,y)$ is given on the area: $0 \le x \le l$,

$0 \le y \le h$, and boundary conditions are also given:

$$f(0,y) = f_1(y),$$

$$f(l,y) = f_2(y),$$

$$f(x,0) = f_3(x),$$

$$f(x,h) = f_4(x),$$

then it's solution is the following expression [3]:

$$f(x,y) = \sum_{n=1}^{\infty} A_n sh\left(\frac{n\pi(l-x)}{h}\right)\sin\left(\frac{n\pi y}{h}\right) + \sum_{n=1}^{\infty} B_n sh\left(\frac{n\pi x}{h}\right)\sin\left(\frac{n\pi y}{h}\right) +$$

$$+ \sum_{n=1}^{\infty} C_n \sin\left(\frac{n\pi x}{l}\right)sh\left(\frac{n\pi(h-y)}{l}\right) + \sum_{n=1}^{\infty} D_n \sin\left(\frac{n\pi x}{l}\right)sh\left(\frac{n\pi y}{l}\right).$$

Coefficients $A_n$, $B_n$, $C_n$, $D_n$ are defined by formulas [3]:

$$A_n = \frac{2}{\lambda_n}\int_0^h f_1(\xi)\sin\left(\frac{n\pi\xi}{h}\right)d\xi,$$

$$B_n = \frac{2}{\lambda_n}\int_0^h f_2(\xi)\sin\left(\frac{n\pi\xi}{h}\right)d\xi,$$

$$C_n = \frac{2}{\mu_n}\int_0^l f_3(\xi)\sin\left(\frac{n\pi\xi}{l}\right)d\xi,$$

$$D_n = \frac{2}{\mu_n}\int_0^l f_4(\xi)\sin\left(\frac{n\pi\xi}{l}\right)d\xi,$$

where $\lambda_n = hsh\left(\frac{n\pi l}{h}\right)$, $\mu_n = lsh\left(\frac{n\pi h}{l}\right)$.

Consider the harmonic function (10):

$$f(x,y) = \frac{\sin(y\alpha)}{\sin(h\alpha)}V_h(x),$$

$$f_1(y) = f(0,y) = \frac{\sin(y\alpha)}{\sin(h\alpha)}V_h(0) = \left(\frac{y}{h} + \frac{yh^2 - y^3}{6h}\alpha^2 + O(\alpha^4)\right)V_h(0) =$$

$$= \left(\frac{y}{h} + \frac{yh^2 - y^3}{6h}\frac{\partial^2}{\partial x^2} + ...\right)V_h(0) = \frac{y}{h}V_h(0),$$

$$f_2(y) = f(l,y) = \frac{\sin(y\alpha)}{\sin(h\alpha)}V_h(l) = \left(\frac{y}{h} + \frac{yh^2 - y^3}{6h}\alpha^2 + O(\alpha^4)\right)V_h(l) =$$

$$= \left(\frac{y}{h} + \frac{yh^2 - y^3}{6h}\frac{\partial^2}{\partial x^2} + ...\right)V_h(l) = \frac{y}{h}V_h(l)$$,

$$f_3(x) = f(x,0) = \frac{\sin(0\alpha)}{\sin(h\alpha)}V_h(x) = 0$$,

$$f_4(x) = f(x,h) = \frac{\sin(h\alpha)}{\sin(h\alpha)}V_h(x) = V_h(x)$$.

It's solution is the following expression:

$$f(x,y) = \sum_{n=1}^{\infty} A_n sh\left(\frac{n\pi(l-x)}{h}\right)\sin\left(\frac{n\pi y}{h}\right) + \sum_{n=1}^{\infty} B_n sh\left(\frac{n\pi x}{h}\right)\sin\left(\frac{n\pi y}{h}\right) +$$

$$+ \sum_{n=1}^{\infty} C_n \sin\left(\frac{n\pi x}{l}\right)sh\left(\frac{n\pi(h-y)}{l}\right) + \sum_{n=1}^{\infty} D_n \sin\left(\frac{n\pi x}{l}\right)sh\left(\frac{n\pi y}{l}\right)$$,

where

$$A_n = \frac{2V_h(0)}{\lambda_n}\int_0^h \frac{\xi}{h}\sin\left(\frac{n\pi\xi}{h}\right)d\xi$$, (18)

$$B_n = \frac{2V_h(l)}{\lambda_n}\int_0^h \frac{\xi}{h}\sin\left(\frac{n\pi\xi}{h}\right)d\xi$$, (19)

$$C_n = 0$$, (20)

$$D_n = \frac{2}{\mu_n}\int_0^l V_h(\xi)\sin\left(\frac{n\pi\xi}{l}\right)d\xi$$. (21)

### 3. THE PROBLEM ON PRESSING OF THE FLAT STAMP IN THE RECTANGULAR PLATE

A flat stamp is pressed into the rectangular plate along the boundary plane $y = h$. We assume that the normal displacements of the plate under the stamp are a known function of $x$, and at the other boundary of the plate, at $y = 0$, are equal to zero. In addition, we assume that at the boundaries of the plate $y = 0$ and $y = h$ there are no shearing stresses. This problem has the following boundary conditions [1]:

$$v(0,y) = \frac{1}{G}V(0,y) = v(l,y) = \frac{1}{G}V(l,y) = 0$$,  (22)

$$\sigma_y(0,y) = Y(0,y) = \sigma_y(l,y) = Y(l,y) = 0$$,

$$\sigma_x(0, y) = \sigma_x(l, y) = 0,$$

$$v(x, h) = \frac{1}{G} V_h(x),$$

$$v(x, 0) = \frac{1}{G} V_0(x) = 0,$$

$$\tau_{xy}(x, h) = X_h(x) = \tau_{xy}(x, 0) = X_0(x) = 0.$$

Considering that the boundary condition (22) is valid for any value of $y$, we write the following equality:

$$V(0, y)\big|_{y=h} = V(l, y)\big|_{y=h} = 0.$$

We perform the substitution:

$$V(0, y)\big|_{y=h} = V(0, h) = V_h(0) = 0, \tag{23}$$

$$V(l, y)\big|_{y=h} = V(l, h) = V_h(l) = 0. \tag{24}$$

Substituting (23) and (24) into the expressions for the coefficients (18) – (21), we obtain:

$$A_n = B_n = C_n = 0,$$

$$D_n = \frac{2}{l\, sh\left(\dfrac{n\pi h}{l}\right)} \int_0^l V_h(\xi) \sin\left(\dfrac{n\pi \xi}{l}\right) d\xi.$$

As a result, we get:

$$\frac{\sin(y\alpha)}{\sin(h\alpha)} V_h(x) = \sum_{n=1}^{\infty} \frac{\sin\left(\dfrac{n\pi x}{l}\right) sh\left(\dfrac{n\pi y}{l}\right)}{sh\left(\dfrac{n\pi h}{l}\right)} \frac{2}{l} \int_0^l V_h(\xi) \sin\left(\dfrac{n\pi \xi}{l}\right) d\xi. \tag{25}$$

Using expressions (11) – (17), we find solutions for the rest harmonic functions:

$$\frac{\cos(y\alpha)}{\sin(h\alpha)} V_h(x) = -\sum_{n=1}^{\infty} \frac{ch\left(\dfrac{n\pi y}{l}\right)\cos\left(\dfrac{n\pi x}{l}\right)}{sh\left(\dfrac{n\pi h}{l}\right)} \frac{2}{l} \int_0^l V_h(\xi) \sin\left(\dfrac{n\pi \xi}{l}\right) d\xi, \tag{26}$$

$$\frac{\alpha \sin(y\alpha)}{\sin(h\alpha)} V_h(x) = \sum_{n=1}^{\infty} \frac{n\pi}{l} \frac{\cos\left(\dfrac{n\pi x}{l}\right) sh\left(\dfrac{n\pi y}{l}\right)}{sh\left(\dfrac{n\pi h}{l}\right)} \frac{2}{l} \int_0^l V_h(\xi) \sin\left(\dfrac{n\pi \xi}{l}\right) d\xi, \tag{27}$$

$$\frac{\alpha \cos(y\alpha)}{\sin(h\alpha)} V_h(x) = \sum_{n=1}^{\infty} \frac{n\pi}{l} \frac{\sin\left(\dfrac{n\pi x}{l}\right) ch\left(\dfrac{n\pi y}{l}\right)}{sh\left(\dfrac{n\pi h}{l}\right)} \frac{2}{l} \int_0^l V_h(\xi) \sin\left(\frac{n\pi \xi}{l}\right) d\xi \tag{28}$$

$$\frac{\alpha^2 \sin(y\alpha)}{\sin(h\alpha)} V_h(x) = -\sum_{n=1}^{\infty} \frac{n^2\pi^2}{l^2} \frac{\sin\left(\dfrac{n\pi x}{l}\right) sh\left(\dfrac{n\pi y}{l}\right)}{sh\left(\dfrac{n\pi h}{l}\right)} \frac{2}{l} \int_0^l V_h(\xi) \sin\left(\frac{n\pi \xi}{l}\right) d\xi \tag{29}$$

$$\frac{\alpha^2 \cos(y\alpha)}{\sin(h\alpha)} V_h(x) = \sum_{n=1}^{\infty} \frac{n^2\pi^2}{l^2} \frac{\cos\left(\dfrac{n\pi x}{l}\right) ch\left(\dfrac{n\pi y}{l}\right)}{sh\left(\dfrac{n\pi h}{l}\right)} \frac{2}{l} \int_0^l V_h(\xi) \sin\left(\frac{n\pi \xi}{l}\right) d\xi \tag{30}$$

$$\frac{\alpha \cos(h\alpha) \cos(y\alpha)}{(\sin(h\alpha))^2} V_h(x) =$$

$$= -\sum_{n=1}^{\infty} \frac{n\pi}{l} \frac{ch\left(\dfrac{n\pi y}{l}\right) \cos\left(\dfrac{n\pi x}{l}\right) ch\left(\dfrac{n\pi h}{l}\right)}{\left(sh\left(\dfrac{n\pi h}{l}\right)\right)^2} \frac{2}{l} \int_0^l V_h(\xi) \sin\left(\frac{n\pi \xi}{l}\right) d\xi \tag{31}$$

$$\frac{\alpha^2 \cos(h\alpha) \cos(y\alpha)}{(\sin(h\alpha))^2} V_h(x) =$$

$$= \sum_{n=1}^{\infty} \frac{n^2\pi^2}{l^2} \frac{\sin\left(\dfrac{n\pi x}{l}\right) ch\left(\dfrac{n\pi h}{l}\right) ch\left(\dfrac{n\pi y}{l}\right)}{\left(sh\left(\dfrac{n\pi h}{l}\right)\right)^2} \frac{2}{l} \int_0^l V_h(\xi) \sin\left(\frac{n\pi \xi}{l}\right) d\xi \tag{32}$$

Substituting $(25) - (32)$ into $(5) - (9)$ we find the exact solution of problem [1]:

$$U(x,y) = -\sum_{n=1}^{\infty} \frac{\delta_n \cos(\beta_n \xi)}{2\Delta_n} \left( \left( (1-2\nu) sh(\beta_n) - \beta_n ch(\beta_n) \right) ch(\beta_n \eta) + \beta_n sh(\beta_n) \eta sh(\beta_n \eta) \right),$$

$$V(x,y) = \sum_{n=1}^{\infty} \frac{\delta_n \sin(\beta_n \xi)}{2\Delta_n} \left( \left( 2(1-\nu) sh(\beta_n) + \beta_n ch(\beta_n) \right) sh(\beta_n \eta) - \beta_n sh(\beta_n) \eta ch(\beta_n \eta) \right),$$

$$Y(x,y) = \sum_{n=1}^{\infty} \frac{\delta_n \beta_n \sin(\beta_n \xi)}{h\Delta_n} \left( \left( sh(\beta_n) + \beta_n ch(\beta_n) \right) ch(\beta_n \eta) - \beta_n sh(\beta_n) \eta sh(\beta_n \eta) \right),$$

$$X(x,y) = \sum_{n=1}^{\infty} \frac{\delta_n \beta_n^2 \cos(\beta_n \xi)}{h\Delta_n} \left( ch(\beta_n) sh(\beta_n \eta) - sh(\beta_n) \eta sh(\beta_n \eta) \right),$$

$$\sigma_x(x,y) = \sum_{n=1}^{\infty} \frac{\delta_n \beta_n \sin(\beta_n \xi)}{h \Delta_n} \Big( \big( sh(\beta_n) - \beta_n ch(\beta_n) \big) ch(\beta_n \eta) + \beta_n sh(\beta_n) \eta\, sh(\beta_n \eta) \Big),$$

where $\eta = \dfrac{y}{h}$, $\xi = \dfrac{x}{h}$, $\Delta_n = (1-\nu)sh^2(\beta_n)$, $\beta_n = \dfrac{n\pi h}{l}$.

Vlasov V.Z. came to this solution in an entirely different way by operating on the series.

## CONCLUSIONS

Usage of the symbolic method and the replacement of harmonic functions by corresponding solutions make it possible to obtain an exact solution of the problem [4,5]. The efficiency of the method is shown on the example of the solution of the problem of pressing a flat stamp into a rectangular plate. Replacing the harmonic functions represented in a symbolic form by the corresponding exact solutions of the harmonic equation for the rectangle, we get the exact solution of the problem.